\documentclass[11pt,twoside]{article}
\usepackage{amsmath}	
\usepackage{asp2010}
\usepackage{natbib}

\def\alf{Alfv\'en}

\usepackage{color}

\def\revise#1{{#1}}

\resetcounters

\begin{document}

\title{An anisotropic-{\alf}ic-turbulence-based solar wind model with proton temperature anisotropy}
\author{Bo Li$^1$ and Shadia Rifai Habbal$^2$}

\affil{$^1$Shandong Provincial Key Laboratory of Optical Astronomy \& Solar-Terrestrial Environment,
   School of Space Science and Physics,
   Shandong University at Weihai, Weihai 264209, China}
\affil{$^2$Institute for Astronomy, University of Hawaii, 2680 Woodlawn Drive, Honolulu, HI 96822, USA}

\begin{abstract}
How the solar wind is accelerated to its supersonic speed is intimately related to how it is heated.
Mechanisms based on ion-cyclotron resonance have been successful in explaining a large number of observations,
   those concerning the significant ion temperature anisotropy above coronal holes in particular.
However, they suffer from the inconsistency with turbulence theory which says that the turbulent cascade
   in a low-beta medium like the solar corona should proceed in the perpendicular rather than the parallel
   direction, meaning that there is little energy in the ion gyro-frequency range for ions to absorb via
   ion-cyclotron resonance.
Recently a mechanism based on the interaction between the solar wind particles and the anisotropic turbulence
   has been proposed, where the perpendicular proton energy addition is via the stochastic heating (Chandran et al. 2011).
We extend this promising mechanism by properly accounting for the effect of proton temperature anisotropy on the
   propagation of \alf\ waves, for the radiative losses of electron energy, and for the field line curvature that
   naturally accompanies solar winds in the corona.
While this mechanism was shown in previous studies to apply to the polar fast solar wind,
   we demonstrate here for the first time that it applies also to the slow wind flowing
   along field lines bordering streamer helmets.
\end{abstract}

\section{Modeling the solar wind: why temperature anisotropy?}
How the solar wind is accelerated is closely related to how it is heated~\citep[e.g.,][]{2009LRSP....6....3C}.
For the nascent fast wind above coronal holes, much evidence exists that ions are hotter than electrons,
    and their perpendicular temperatures are higher than the parallel one ($T^{\perp}>T^{\parallel}$)
    (see data compiled in \citeauthor{2002SSRv..101..229C}~\citeyear{2002SSRv..101..229C},
    also \citeauthor{2008ApJ...678.1480C}~\citeyear{2008ApJ...678.1480C}).
Interestingly, in the near-Sun region the slow wind flanking streamer helmets exhibits a similar tendency as evidenced by
    measurements with SOHO/UVCS~\citep{2002ApJ...571.1008S,2003ApJ...597.1145F}.
Given that the solar winds receive most acceleration close to the Sun,
    these observations point to
    the possibility that the fast and slow winds may share a common heating, and hence acceleration, mechanism.
Some other factors, most notably the distinct flow tube geometry, then determine the heating profile and
    hence lead to significantly
    different terminal speeds of the two winds~\citep{1990ApJ...355..726W, 2007ApJS..171..520C}.

The temperature measurements, especially the inferred significant ion temperature anisotropy,
    lead naturally to the suggestion that the nascent solar winds are heated by ion-cyclotron
     resonance~\citep[e.g.,][]{2002JGRA..107.1147H}.
The needed high frequency ion-cyclotron waves (ICWs) may be generated either by a turbulent parallel cascade from low-frequency waves
    emitted by the Sun, or directly by small-scale magnetic reconnection events at chromospheric network.
While the latter remains safe, the former has been challenged recently as both MHD tubulence theory and
    simulation studies~\citep[e.g.,][]{2003LNP...614...56C}
    suggest that in a low-$\beta$ solar corona, the turbulence cascade prefers the $\perp$-direction, hence generating modes with high $k_\perp$
    instead of ICWs with high $k_\parallel$ and high frequency.
However, these high-$k_\perp$ modes, presumably kinetic \alf\ waves (KAWs), are most readily dissipated by electrons
    in a low-$\beta$ plasma, if the only dissipation channel is via the Landau resonance.
The consequent preferential electron heating would violate available measurements.
A remedy comes only recently, where KAWs with sufficient amplitudes were demonstrated to primarily heat
    protons, preferentially in the $\perp$-direction, by rendering the
      ion orbits stochastic~\citep{2010ApJ...720..503C}.
This nonlinear mechanism, termed ``stochastic heating'', was further parameterized~\citep{2010ApJ...720..548C} and
     eventually incorporated into a two-fluid
    solar wind model~\citep{2011ApJ...743..197C}(hereafter C11) where the anistropic turbulence was shown to be capable of producing
    polar fast solar wind solutions that agree with a substantial number of observations.

Given that the fast and slow winds may share a common heating mechanism, and that C11 is restricted to
    the polar fast wind, we wish to extend C11 by asking whether the mechanism therein also applies to
    the equatorial slow wind as well.
To address this, we construct slow wind models that are based on the anisotropic turbulence, that disinguish between
    electrons and protons, and that account for the proton temperature anisotropy.
Inherited from C11 are:
    1) The turbulence is driven by reflections of outward ($z^+$) waves off the gradients of the solar wind parameters,
    2) Outward waves $z^+$ dominate inward ones $z^-$,
    3) $z^-$ waves are cascaded sufficiently fast compared with their wave periods,
    4) the cascade proceeds in the inertial range along a ``critical balance'' path until it reaches a $\perp$-scale comparable to the ion gyroradius,
       where the generated KAWs are damped via both Landau damping (yielding electron heating, and proton $\parallel$-heating),
       and stochastic heating (yielding proton $\perp$-heating),
    5) if the wave energy is cascaded to smaller still $\perp$-scales, only electrons receive this energy.
However, compared with C11, we:
    1) properly incorporate in the gross dissipation rate the field line curvature that accompanies
       field lines bordering streamer helmets,
    2) incorporate the radiative loss in the electron energy equation, which is important in the energy balance
         in the upper transition region,
    3) improve the treatment of wave propagation by including the effects of temperature anisotropy, which was missing in the original treatment in C11.
{\revise{For simplicity, instead of a more self-consistent, collisionless treatment of the proton heat fluxes based on the Landau fluid approach as was adopted by C11,
    we assume that the fluxes can be described by a modified Spitzer form.}}
In what follows, we describe the improved model in section~\ref{sec_model},
    present a constructed slow wind model in section~\ref{sec_model_res},
    and summarize our effort in section~\ref{sec_disc}.

\section{Model desciption and method of solution}
\label{sec_model}
The fluid part of our model starts with the standard 16-moment transport equations
   (see appendix in~\citeauthor{2009A&A...494..361L}~\citeyear{2009A&A...494..361L}).
When axial symmetry is assumed and solar rotation neglected, in a steady state the vector equations
     may be decomposed
    into a force balance condition across the meridional magnetic field lines
    and a set of transport equations along them.
For simplicity we replace the former by prescribing a meridional magnetic field configuration
    presented in~\citet{1998A&A...337..940B}, which is representative of a minimum corona.
The transport equations along curved field lines then read
\begin{align}
& \left(n v a\right)' =0, 	\label{eq_density} \\
& vv'       +\frac{k_B}{ n m_p}\left[n \left(T_p^{\parallel}+T_e\right)\right]'
            +\frac{k_B(T_p^{\parallel}-T_p^\perp)a'}{a m_p}
            - \left(\frac{G M_\odot}{r}\right)' -\frac{F}{n m_p} = 0,
	\label{eq_momen} \\
& v \left(T_e\right)'
         + \frac{(\gamma-1)T_e \left(a v\right)' }{a}
         -\frac{\gamma-1}{n k_B a}\left(a \kappa_{e0} T_e^{5/2} T_e'\right)' \nonumber  \\
& \qquad + 2\nu_{pe} (T_e-T_p) - \frac{\gamma-1}{n k_B} (Q_e - L_{\mathrm{rad}}) = 0, \label{eq_Te} \\
& v (T_p^\parallel)'
         + 3(\gamma-1)T_p^\parallel v'
         -\frac{\gamma-1}{n k_B a}\left[a \tilde{\kappa}_{p 0} T_p^{5/2} \left(T_p^{\parallel}\right)'\right]' \nonumber  \\
& \qquad + 2\nu_{pe} (T_p^\parallel-T_e) + 2\nu_{pp} (T_p^\parallel-T_p^{\perp})
- \frac{3(\gamma-1) Q_p^\parallel}{n k_B} = 0, \label{eq_tppara} \\
& v (T_p^\perp)'
         + \frac{3 (\gamma-1) v T_p^\perp a' }{2 a}
         - \frac{\gamma-1}{ n k_B a}\left[a \tilde{\kappa}_{p 0} T_p^{5/2} \left(T_p^{\perp}\right)'\right]' \nonumber  \\
& \qquad + 2\nu_{pe} (T_p^{\perp}-T_e) + \nu_{pp} (T_p^{\perp}-T_p^\parallel)
         - \frac{3(\gamma-1) Q_p^\perp}{2n k_B}  = 0. \label{eq_tpperp}
\end{align}
Here $n$ is the number density and $v$ the speed, $a\propto 1/B$ is the tube cross-sectional area
     with $B$ being the magnetic field strength.
An arbitrary point along a field line is characterized by both $(r,\theta)$ (its heliocentric distance and colatitude) and
     arclength $l$.
The prime  $'$ denotes the differentiation with respect to $l$.
Moreover, $k_B$ and $m_p$ are the Boltzmann constant and proton mass.
The gravitational constant is denoted by $G$, and solar mass by $M_\odot$.
The electron, proton parallel and perpendicular temperatures are denoted by $T_e$, $T_p^\parallel$
    and $T_p^\perp$, respectively.
The mean proton temperature is given by $T_p = (T_p^\parallel+ 2T_p^\perp)/3$.
Furthermore, $\gamma=5/3$ is the adiabatic index.
The Coulomb collision rates $\nu_{pe}$ and $\nu_{pp}$
    are evaluated by using a Coulomb logarithm of $23$.
For simplicity, the Spitzer law is assumed for the electron heat flux, $\kappa_{e0} = 7.8\times 10^{-7}$
    (cgs units will be used throughout).
A similar form is assumed for the proton ones, but the coefficient is arbitrarily reduced relative to
    the Spitzer value $\kappa_{p, c} = 3.2\times 10^{-8}$ by a spatially varying factor,
    namely, $\tilde{\kappa}_{p0}/\kappa_{p, c}$ is $1$ for $r<3$~$R_\odot$,
    ramped linearly with $r$ to $0.02$ at $r=10$~$R_\odot$ and kept so
    from there on~\citep[see e.g.,][]{1999JGR...104.2521L}.
Besides, $L_{\mathrm{rad}}$ represents the radiative losses and we adopt the standard
    parametrization by~\citet{1978ApJ...220..643R} for an optically thin medium.
$Q_e$, $Q_p^{\parallel}$ and $Q_p^\perp$ are the heating rates, while $F$ represents the volumetric
    force density the solar wind receives.
Part of the electron heating is assumed to come from some ad hoc process which operates close to the base,
    $Q_{e, \mathrm{basal}} = Q_{e0}\exp(-l/l_d)$.
This basal heating is found to be important to improve the agreement of the model results with observations.
On the other hand, the majority of the electron heating $Q_{e,\mathrm{wav}}$, as well as
    the proton heating rates, are due entirely to the turbulent dissipation of low-frequency \alf\ waves.

\begin{figure}
\centering
\includegraphics[width=0.9\textwidth]{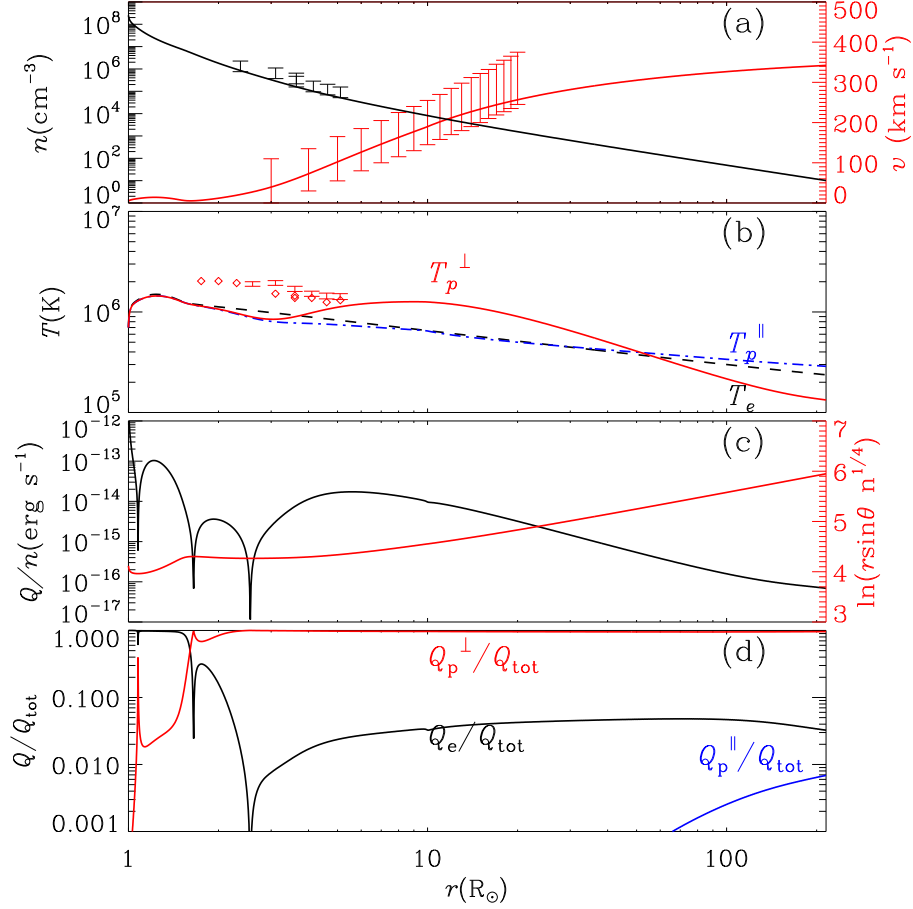}
\caption{
Radial distribution of solar wind parameters derived from an anistropic-turbulence-heated 2-fluid model
    with proton temperature anisotropy.
(a): number density $n$ and speed $v$.
(b): electron, proton parallel and perpendicular temperatures $T_e$, $T_p^\parallel$, and $T_p^\perp$.
(c): gross turbulence dissipation rate per particle $Q_{\mathrm{wav}}/n$ and the geometrical factor $\ln(r\sin\theta n^{1/4})$.
(d): fractions of $Q_{\mathrm{wav}}$ that goes to heating electrons and protons.
In (a) and (b), the error bars and open boxes represent some remote-sensing measurements representative
    of solar minimum conditions (please see text).
}
\label{fig_slowSW}
\end{figure}

The wave part of the model is a WKB evolution equation with dissipation terms,
\begin{align}
 \frac{\left(a F_w\right)'}{a} + v F = -Q_{\mathrm{wav}} . \label{eq_wave}
\end{align}
Here only the outward waves contribute to the wave energy flux density $F_w$ and wave force $F$, whose expressions are given by
    Eqs.(18) and (16) in~\citet{1999JGR...104.2521L},
    where the correction due to finite temperature anisotropy is included.
On the other hand, the gross dissipation rate does take into account the inward waves, and the interaction between outward and
    inward ones yields
\begin{align}
 Q_{\mathrm{wav}} = 4 c_d p_w (v+v_A) \left|\left[\ln\left( r \sin\theta n^{1/4}\right) \right]'\right| , \label{eq_qwav}
\end{align}
    where $p_w$ is the wave pressure, $v_A$ the \alf\ speed, and $c_d$ is a dimensionless constant.
Note that the field line shape appears here by contributing further reflection via the gradient of the geometrical factor
    $\sin\theta$.
To apportion $Q_{\mathrm{wav}}$ among $Q_{e, \mathrm{wav}}$ and $Q_{p}^{\parallel, \perp}$, Eqs.(44) to (56) in C11 are used.
It suffices to mention here that a correlation scale $L_\perp \propto 1/\sqrt{B}$ appears not directly in $Q_{\mathrm{wav}}$,
   as in ICW-based models~\citep{2002JGRA..107.1147H}, but in the way $Q_{\mathrm{wav}}$ is apportioned.

To proceed, the field line flanking some streamer helmet is characterized by the $B$-distribution
   and its shape.
These are imposed by requiring that the field line reach $88^\circ$ colatitude at 1~AU in the~\citet{1998A&A...337..940B} model
   with their fiducial parameters {\revise{(the DQCS model as presented in Fig.3 therein)}}, and that $B$ at 1~AU be 3.5~$\gamma$.
The wind is determined by the following set of parameters: the base density $n_\odot$, temperature $T_\odot$, and
    wave amplitude $\delta v_\odot$, as well as $L_\perp$ at the base $L_\odot$.
We choose $n_\odot = 2.5\times 10^8$~cm$^{-3}$, $T_\odot = 7\times 10^5$~K,
    $\delta v_\odot = 35$~km~s$^{-1}$, and $L_\odot=100$~km.
Once these are prescribed, {\revise{we first cast equations~(\ref{eq_density})
    to (\ref{eq_wave}) in a time-dependent form and then evolve them
    from an arbitrary initial state}}  until a steady state is found, which is then examined in some detail in next section.

\section{Constructed slow solar wind model}
\label{sec_model_res}

Figure~\ref{fig_slowSW} presents the radial distribution between 1~$R_\odot$ and 1~AU of
    a number of solar wind parameters.
In Fig.\ref{fig_slowSW}a, the computed number density $n$ is compared with the electron density measurements
     presented in~\citet{2002ApJ...571.1008S} (their {fig.3c}),
    while the speed profile $v$ is compared with the range of wind speeds derived by
     tracking a collection of small inhomogeneities (the blobs) in
     images obtained with SOHO/LASCO~\citep{2000JGR...10525133W}.
In addition to obtaining reasonable values of speed $v=342$~km~s$^{-1}$ and flux density
     $n v = 3.5\times 10^8$~cm$^{-2}$~s$^{-1}$
     at 1~AU,
     the model results agree reasonaly well with the observations close to the Sun.
The temperatue profiles, presented in Fig.\ref{fig_slowSW}b, are found to be difficult to
     reproduce the measured H I ones obtained with SOHO/UVCS and
     represented by the open boxes~\citep{2002ApJ...571.1008S} as well as error bars~\citep{2003ApJ...597.1145F}.
Despite this, the resultant pressure gradient force produces a decent slow wind solution.
It is interesting to see that, the gross turbulence dissipation rate $Q_{\mathrm{wav}}$ exhibits
     multiple zeroes (Fig.\ref{fig_slowSW}c, black curve), as a result of
     the geometrical factor $\ln(r\sin\theta n^{1/4})$ (Fig.\ref{fig_slowSW}c, red curve) attaining its local extrema.
This is understandable given that $Q_{\mathrm{wav}}$ is proportional to the gradient
     of this factor (Eq.\ref{eq_qwav}).
Also noteworthy is that, among the dissipation channels of the wave energy,
     the Landau damping of KAWs does not make any significant contribution (Fig.\ref{fig_slowSW}d).
When the electron heating dominates ($r<1.6$~$R_\odot$), the dissipation is due to the processes
     that take place at
     perpendicular scales shorter than the proton gyroradius.
On the other hand, when the proton heating dominates ($r>1.6$~$R_\odot$), the dissipation is almost entirely due to
     stochastic heating of KAWs, evidenced by the fact that protons literally receive no parallel heating.
{\revise{At distances $r\gtrsim 122$~$R_\odot$ where $T_p^{\perp}/T_p^{\parallel}$ is substantially smaller than unity,
    the solution becomes firehose unstable. However, this should have little effect on the gross wind parameters given that
    the wind itself has already been fully accelerated.
      }}

\section{Summary}
\label{sec_disc}
So far the rather promising mechanism, originated by~\citet{1999ApJ...523L..93M} and further developed
     by, to name but a few, \citet{2005A&A...444..233V, 2007ApJ...662..669V, 2010ApJ...708L.116V, 2007ApJS..171..520C, 2012ApJ...754...92C, 2011ApJ...743..197C}, is almost entirely devoted to
     examining the polar fast solar wind.
Does it also apply to the slow wind that flows along curved magnetic field lines bordering streamer helmets?
Our preliminary results suggest that this is indeed the case.

\acknowledgements
This research is supported by the National Natural Science Foundation of China (40904047 and 41174154), the Ministry of Education of China
     (20110131110058 and NCET-11-0305).

\bibliographystyle{asp2010}
\bibliography{LiHabbal}

\end{document}